\documentclass[12pt]{article}

\usepackage{amssymb,amsmath,amsfonts,eurosym,geometry,ulem,graphicx,caption,color,setspace,sectsty,comment,footmisc,caption,pdflscape,subfigure,array,hyperref}
\usepackage[semicolon]{natbib}
\usepackage{amsmath}				% For Math
\usepackage{fancyhdr}				% For fancy header/footer
\usepackage{graphicx}				% For including figure/image
\usepackage{cancel}					% To use the slash to cancel out stuff in work
\usepackage{amsmath}
\usepackage{amssymb}
\usepackage{geometry}
\usepackage{float}
\usepackage{listings}
\usepackage{adjustbox}
\usepackage{float}
\usepackage{hyperref} 
\usepackage{newfloat}

\DeclareFloatingEnvironment[
    fileext=loc,
    listname={List of Code Listings},
    name=Code,
    placement=H,
    within=section,
]{codelisting}

\newcolumntype{L}[1]{>{\raggedright\let\newline\\arraybackslash\hspace{0pt}}m{#1}}
\newcolumntype{C}[1]{>{\centering\let\newline\\arraybackslash\hspace{0pt}}m{#1}}
\newcolumntype{R}[1]{>{\raggedleft\let\newline\\arraybackslash\hspace{0pt}}m{#1}}

\begin{document}

\begin{titlepage}
\title{Blending Ensemble for Classification \\ \large with Genetic-algorithm generated Alpha factors and Sentiments (GAS)}
\author{Quechen Yang}
\date{\today}
\maketitle
\begin{abstract}
\noindent With the increasing maturity and expansion of the cryptocurrency market, understanding and predicting its price fluctuations has become an important issue in the field of financial engineering. This article introduces an innovative Genetic Algorithm-generated Alpha Sentiment (GAS) blending ensemble model specifically designed to predict Bitcoin market trends. The model integrates advanced ensemble learning methods, feature selection algorithms, and in-depth sentiment analysis to effectively capture the complexity and variability of daily Bitcoin trading data.
The GAS framework combines 34 Alpha factors with 8 news economic sentiment factors to provide deep insights into Bitcoin price fluctuations by accurately analyzing market sentiment and technical indicators. The core of this study is using a stacked model (including LightGBM, XGBoost, and Random Forest Classifier) for trend prediction which demonstrates excellent performance in traditional buy-and-hold strategies.
In addition, this article also explores the effectiveness of using genetic algorithms to automate alpha factor construction as well as enhancing predictive models through sentiment analysis. Experimental results show that the GAS model performs competitively in daily Bitcoin trend prediction especially when analyzing highly volatile financial assets with rich data.
\vspace{0in}
\vspace{0in}\\

\bigskip
\end{abstract}
\setcounter{page}{0}
\thispagestyle{empty}
\end{titlepage}
\pagebreak \newpage
\setcounter{page}{0}
\tableofcontents
\newpage

\doublespacing
\section{Introduction} \label{sec:introduction}
Cryptocurrencies have become a very popular topic, mainly due to their disruptive potential and unprecedented returns. Cryptocurrencies are digital currencies that use blockchain technology, a decentralized cryptographic technology capable of facilitating trust digitally. In 2017 and early 2018, people's frenzied interest in cryptocurrency was mainly caused by news reports of unprecedented returns on investment in cryptocurrencies, which then attracted a gold rush. Due to the difficulty for investors to judge whether the information released is true or false, this effect has been further strengthened. As the cryptocurrency market is relatively young, traditional news media does not always report events in a timely manner, leading social media to become the main source of information for cryptocurrency investors. \cite{kraaijeveld2020predictive} demonstrated a dynamic relationship between Twitter sentiment and cryptocurrency prices using bivariate Granger causality.
\\
\\
Except for the sentiment factors, Bitcoin generates abundant trading data due to its high liquidity everyday. Considering the multitude of variables influencing these changes, it's exceedingly challenging to identify the most suitable variables, methods, and parameters for predicting price changes.
 \cite{mcnally2018predicting}  used an ensemble of RNN and LSTM for predicting Bitcoin prices, optimizing it with the Boruta algorithm for feature engineering, which functions similarly to the random forest classifier. 
\cite{hasan2022blending} employed a novel blending ensemble learning model (LKDSR) for predicting crude oil prices. This method combines various machine learning approaches, including linear regression, k-nearest neighbors regression, regression trees, support vector regression, and ridge regression. It focused on short-term imbalances in oil prices caused by multiple frequency factors. The effectiveness of this model in various short-term crude oil price fluctuation patterns indicates its potential applicability in other financial and commodity markets.
\cite{wang2022stock} explored the application of Gradient Boosting Machine (GBM) and its variants, like Extreme Gradient Boosting (XGB) and Light Gradient Boosting Machine (LGBM), in financial data analysis. They concluded that despite deep learning models (like CNNs) performing well in multiple domains, GBM and its variants often show better performance with tabular data, especially in terms of interpretability, quick convergence, and integrating domain knowledge. 
\cite{nti2020comprehensive} evaluated the most suitable ensemble methods in stock market forecasting, including bagging, boosting, stacking, and blending. Using data from stock market indices of four countries (GSE, BSE, NYSE, and JSE), the results indicated that the stacking technique used to build ensemble classifiers or regressors performed best among all tested techniques. They also proposed that future work should include using methods like Genetic Algorithms (GA) to optimize feature selection and parameter settings for SVMs. 
\cite{sivri2024adaptive} introduced an LGBM-SHAP framework, combining the SHAP feature selection method that consistently delivered superior results among various approaches and the LGB's outstanding predictive ability in terms of profitability and risk-adjusted returns. This was applied to 30 different stocks, with results significantly outperforming the BIST 30 index.
\\
\\
This paper proposes a Genetic-algorithm generated Alpha factors Sentiment (GAS) blending ensemble model for Bitcoin. The framework combines ensemble learning and feature selection algorithms to effectively capture the daily trend of Bitcoin. The focus of this study is to predict the trend of the second day's return rate. The framework consists of thirty-four alpha factors and eight news economic sentiment factors. Experimental results show that the prediction by a stacking model using LGBM, XGB and Random Forest Classifier (RFC) has achieved competitive performance compared to buy-and-hold strategy.

\section{Experiment} \label{sec:experiment}

This experiment is conducted under one 3090 50GB GPU, with Tensorflow, CUDA and Python versions being 2.5.0, 11.2, 3.8 respectively.
The run-time of sentiment classification is around 5 hours and primarily depends on the evaluation frequency, requiring recalculating embedding for all news.
And blending ensemble model has a run-time of approximately 71 hours if applying the grid search method for all three base learners.

\paragraph{Datasets}
The financial indicator dataset used in this experiment was downloaded from Yahoo Finance, including financial indicators such as open, high, low, close, Adj Close and Volume. After datasets warrant cleaning and data imputation (using the ffill method to prevent Data Leakage, that is, knowing the next period's factor level in advance), a total of 3109 daily trading day data from 14/07/2015 to 16/01/2024 were retained.\\
\\
The news heatmap data used in this experiment consists of a total of 377,295 records, covering a total of 1,354 daily trading day data from 10/10/2019 to 13/07/2023. The collection method for this data comes from the daily news crawl of my last internship. The data source is \url{https://www.jin10.com/}, China's largest trading platform. I translated every news using Google translate API and input it into a transformer model. For specific details on the classification method for the news, refer to section 2.1.1.
\\
\\
Given that the algorithm needs to be retrained every day, I pay special attention to avoiding information loss caused by daily data updates. Therefore, the prediction period set in the experiment is one day. To address this setting, I used time series cross-validation (TSCV) method. Compared with traditional sliding window methods, TSCV is more suitable for handling data with time dependence and adapts to dynamic changes in daily markets while preventing information loss. Under the framework of TSCV, the model regularly receives updates of the latest data to ensure its predictive accuracy and relevance. In this process, data is divided into multiple continuous time windows instead of random partitions to simulate actual time flow and market changes. I ensure that each subset of training and testing data is continuous.
\\
\\
A threshold of $0.1 \%$ is set for extremely small near-zero returns, utilizing the corresponding $\log$ returns as the dependent variable. This approach ensures a balanced training set with ample negative examples. Labels are designated as 0 or 1 based on the comparison of log returns to the threshold.
% 111111111111111
\begin{figure}[H]
    \centering
    \fbox{\includegraphics[scale=0.5]{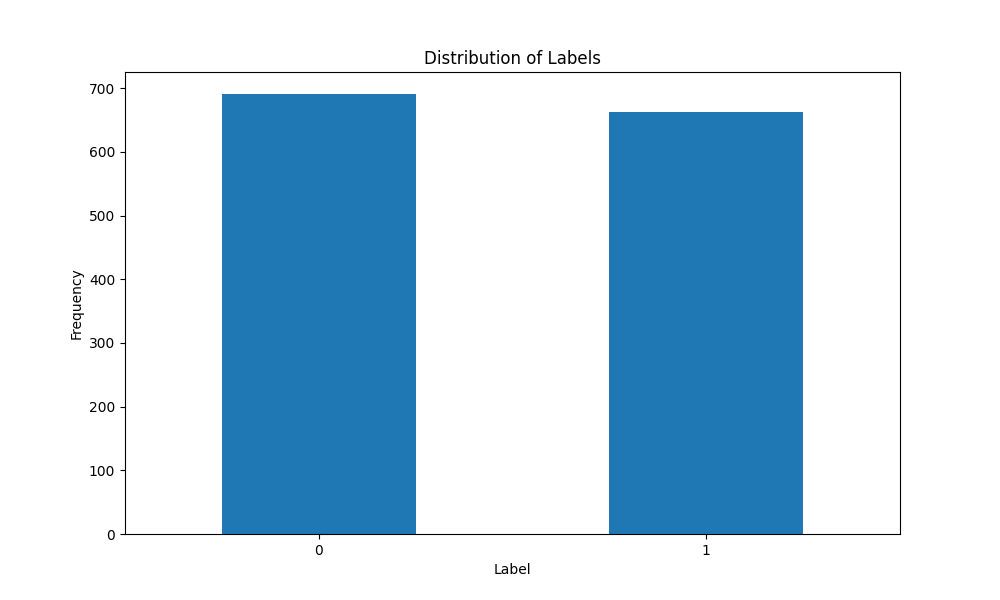}} 
    \caption{labels distribution of the datasets}
    \label{fig:mc_ver}
\end{figure}

\subsection{Feature Engineering} 

\subsubsection{Feature Creation}
Technical indicators such as Moving Averages (MA), Kaufman's Adaptive Moving Average (KAMA), and Moving Average Convergence Divergence (MACD) are used for tracking and smoothing market trends. Others, like the Stochastic Oscillator (KDJ) and Relative Strength Index (RSI), measure market momentum and strength from different perspectives. These indicators often signal potential price reversals when reaching extreme high or low points, and can be constructed into factors that reflect the medium to long-term trends in stock prices. Effectively filtering out market "noise," they help investors identify and follow market trends. Below are the technical indicator factors I use:
\\
\begin{itemize}
  \item \textbf{Trend Following Factors:}
  \begin{itemize}
    \item Difference between price and Moving Averages: Signifies potential trend reversals when price significantly deviates from its moving averages.
    \item \textbf{Moving Average (MA)}: Calculated by averaging stock prices over a specific period, used to smooth price fluctuations and identify trends.
    \item \textbf{Kaufman's Adaptive Moving Average (KAMA)}: A variable period moving average that adapts to changes in market volatility.
    \item \textbf{Moving Average Convergence Divergence (MACD)}: An indicator showing the difference between two moving averages, commonly used to identify trend changes.
  \end{itemize}

  \item \textbf{Baseline Features - Trend/Volatility Factors:}
  \begin{itemize}
    \item Key features extracted from stock market data, such as the Open-Close price difference (O-C), High-Low price difference (H-L), and GAP, help capture market volatility and price dynamics. These basic price differences reveal market sentiment and potential trend reversal signals, useful for identifying high-risk market areas and potential trading opportunities.
  \end{itemize}

  \item \textbf{Reversal Signal Factors:}
  \begin{itemize}
    \item Factors predicting market reversals are created by analyzing the deviation of prices from their historical performance, such as lags and increments. These factors are crucial for capturing market tops and bottoms.
    \item \textbf{KDJ (Stochastic Oscillator)}: Analyzes price momentum, helping identify overbought or oversold conditions.
    \item \textbf{RSI (Relative Strength Index)}: Combined with ranking and proportional scale functions, factors are constructed to display the relative strength or weakness of stocks against the market, identifying market leaders and laggards.
  \end{itemize}

  \item \textbf{Other Factors:}
  \begin{itemize}
    \item \textbf{Bollinger Bands (BOLL)}: Consisting of an upper, middle, and lower band, it measures price volatility and potential buy-sell ranges.
    \item \textbf{Commodity Channel Index (CCI)}: Measures the degree of deviation of stock prices from their statistical average, used to identify extreme market conditions.
    \item \textbf{ROC (Rate of Change)} and \textbf{ROCP (Rate of Change Percentage)}: Measures the rate of price change, aiding in identifying market momentum.
    \item \textbf{MOM (Momentum)}: Measures the magnitude of price changes, reflecting market momentum.
    \item \textbf{ATR (Average True Range)}: Measures market volatility, used in risk management.
    \item \textbf{WILLR (Williams \%R)}: Positions the current price relative to its high-low range over a specified past period.
  \end{itemize}
\end{itemize}
The formulas constructed as follows:

\begin{table}[H]
\begin{adjustbox}{width=\textwidth,center}
\begin{tabular}{ll}
\hline
\textbf{Indicator}                        & \textbf{Formula}                                                                                                     \\ \hline
Moving Average (MA)                       & $ \text{MA}_n = \frac{\sum_{i=1}^{n} \text{Close}_i}{n} $                                                             \\
Kaufman's Adaptive Moving Average (KAMA)  & Implemented using TA-Lib library                                                                                     \\
MACD                                      & $ \text{MACD} = \text{EMA}_{\text{short}} - \text{EMA}_{\text{long}} $                                                \\
Open-Close Difference (O-C)               & $ \text{O-C} = \text{Open} - \text{Close} $                                                                           \\
High-Low Difference (H-L)                 & $ \text{H-L} = \text{High} - \text{Low} $                                                                             \\
GAP                                       & $ \text{GAP} = \frac{\text{Today's Open}}{\text{Previous Close}} - 1 $                                                \\
Cumulative Log Return (CumLgReturn)       & $ \text{CumLgReturn}_{i\text{d}} = \sum_{t=0}^{i-1} \text{Log Returns}_t $                                            \\
KDJ (Stochastic Oscillator)               & Implemented using TA-Lib library                                                                                        \\
RSI (Relative Strength Index)             & Implemented using TA-Lib library             \\
Bollinger Bands (BOLL)                    & $ \text{Bollinger Bands} = \text{MA} \pm \text{Standard Deviation Multiplier} $                                       \\
Commodity Channel Index (CCI)             & $ \text{CCI} = \frac{\text{Typical Price} - \text{MA of TP}}{0.015 \times \text{Mean Deviation}} $                    \\
Rate of Change (ROC)                      & $ \text{ROC} = \frac{\text{Close}_t - \text{Close}_{t-n}}{\text{Close}_{t-n}} $                                      \\
Rate of Change Percentage (ROCP)          & $ \text{ROCP} = \frac{\text{Close}_t - \text{Close}_{t-n}}{\text{Close}_{t-n}} \times 100\% $                        \\
Momentum (MOM)                            & $ \text{MOM}_t = \text{Close}_t - \text{Close}_{t-n} $                                                                \\
Average True Range (ATR)                  & $ \text{ATR}_t = \text{Average}(\text{High}_t - \text{Low}_t, |\text{High}_t - \text{Close}_{t-1}|, |\text{Low}_t - \text{Close}_{t-1}|) $ \\
Williams \%R (WILLR)                       & $ \text{WILLR}_t = \frac{\text{High}_n - \text{Close}_t}{\text{High}_n - \text{Low}_n} \times -100\% $               \\ \hline
\end{tabular}
\end{adjustbox}
\caption{Financial Indicators and Their Formulas}
\end{table}
As the table mentioned, TA-Lib is a leading technical analysis library that originated from Mario Fortier and has quickly gained popularity. Ta-Lib has over 150 indicators, including ADX, MACD, RSI, Bollinger Bands and candlestick pattern recognition to simplify complex analysis. Instead of writing a few lines of code first to find the upper and lower bands and then the Bollinger band, I use Ta-Lib and input just one line of code to get the result \citep{AdvancedML1}.
\\
By combining basic mathematical operations such as addition, subtraction, multiplication, division, maximum value, minimum value and other advanced statistical features such as ranking, time series minimum value and maximum value, the ability to construct complex and multi-dimensional factors is provided. \citep{AdvancedML1}.

\paragraph{Genetic Algorithm}
In 2016, the quantitative investment management firm, WorldQuant, made public 101 formulaic alpha factors in \cite{kakushadze2016101}. Since then, many quantitative trading methods have used these formulaic alphas for stock trend prediction. 
\\
In quantitative finance, Alpha factors refers to the excess return of an investment relative to a benchmark index. It is used to identify potential profit opportunities that cannot be explained by overall market trends. The higher the value, the greater the likelihood that the stock will have relatively large returns in the following days \citep{zhaofan2022genetic}. 
\\
\\
Artificially constructing Alpha factors is difficult because it is challenging to extract nonlinear factors from high-dimensional and noisy financial data. The most common method of generating new formulas for Alpha is for economists or financial engineers to propose new economic ideas, convert these ideas into formulas, and then verify their effectiveness on historical stock datasets. In order to better adapt to rapidly changing investment markets, more and more scholars are turning towards trading data information such as transaction volume, transaction price, turnover rate and other indicators that can reflect changes in stock prices more accurately with hopes of improving investment efficiency and quality. By using automated factor construction methods for factor mining they hope to gain a more comprehensive understanding and explanation of market excess returns \cite{zhang2020autoalpha}.
\\
\\
The Genetic Algorithm (GA) process of GAS is as follows, with 'gplearn' being a popular Python package that implements standard genetic algorithms.
\begin{itemize}
    \item Generate a batch of Alpha factors as the initial population and calculate the fitness of each factor in the initial population.
    \item The algorithm runs for a set of 5 generations. Each generation's evolution is based on the fitness of the expressions from the previous generation. The population size is fixed at 5000, with a tournament size of 1000, to maintain a balance between exploration and exploitation of the solution space.
    \item In each round of evolution, factors are selected based on their fitness, and new populations are generated through genetic operations such as crossover and mutation. Expressions ranked in the top 1000 by fitness are retained between generations to ensure high-quality solutions are not lost.
    \item Calculate the fitness of each factor in the new population and eliminate some factors based on fitness, ensuring the factors in the population have both high returns and low correlation. The population size is maintained at 350 factors or fewer.
    \item The new population serves as the parent for the next round of evolution.
    \item The iteration ends when it reaches 5 generations or when there is no further improvement in factor fitness, yielding a set of optimally evolved symbolic expressions.
\end{itemize}
In the genetic algorithm, I adopted the 'half and half' initialization method, which combines the 'grow' and 'full' strategies to generate a diversified initial population. In the 'grow' strategy, formulas and base data are randomly drawn, while in the 'full' strategy, leaf nodes draw base data, and non-leaf nodes draw formulas. The 'half and half' strategy combines these two methods, using 'grow' for half and 'full' for the other half, to achieve a better quality of the initial population.
\\
\\
I use the cross-sectional Spearman correlation coefficient between the alpha factor value of the selected stocks and the return rate of the stock in the next period to judge the predictive ability of the factor to calculate the return rate of the stock in the next period, that is the Information Coefficient (IC), which can be calculated by the following formula:
\begin{equation}
    IC = \operatorname{correlation}(R^{T+1}, f) = \frac{\operatorname{Cov}(R^{T+1}, f)}{\operatorname{Std}(R^{T+1}) \operatorname{Std}(f)}
\end{equation}
Where \( R^{T+1} \) is the next period of asset return and alpha factor value \( f \). The larger IC value of an Alpha factor is, the stronger stock selection ability of the factor is.

\paragraph{Sentiment}
In the field of natural language processing (NLP), sentiment analysis aims to identify and classify the emotional tendencies in text. With the latest advances in deep learning, especially the widespread use of Transformer architecture, the accuracy and efficiency of sentiment analysis have been significantly improved. \cite{li2019sentiment} directly used Twitter sentiment as a key indicator to predict price fluctuations for ZClassic, a low market value alternative cryptocurrency. The model's predictions showed strong correlation with historical price data, demonstrating that social media platforms like Twitter can be powerful indicators for predicting speculative trends in alternative cryptocurrency markets. 
\\
\\
Although this subject does not focus on deep learning, I built a specific locally evaluation system for sentiment to improve factor performances of GAS. Since this part is not the main focus of this subject, I will briefly describe it.
% image3-OPRF
\begin{figure}[H]
    \centering
    \begin{minipage}{.27\textwidth}
        \centering
        \fbox{\includegraphics[scale=0.06]{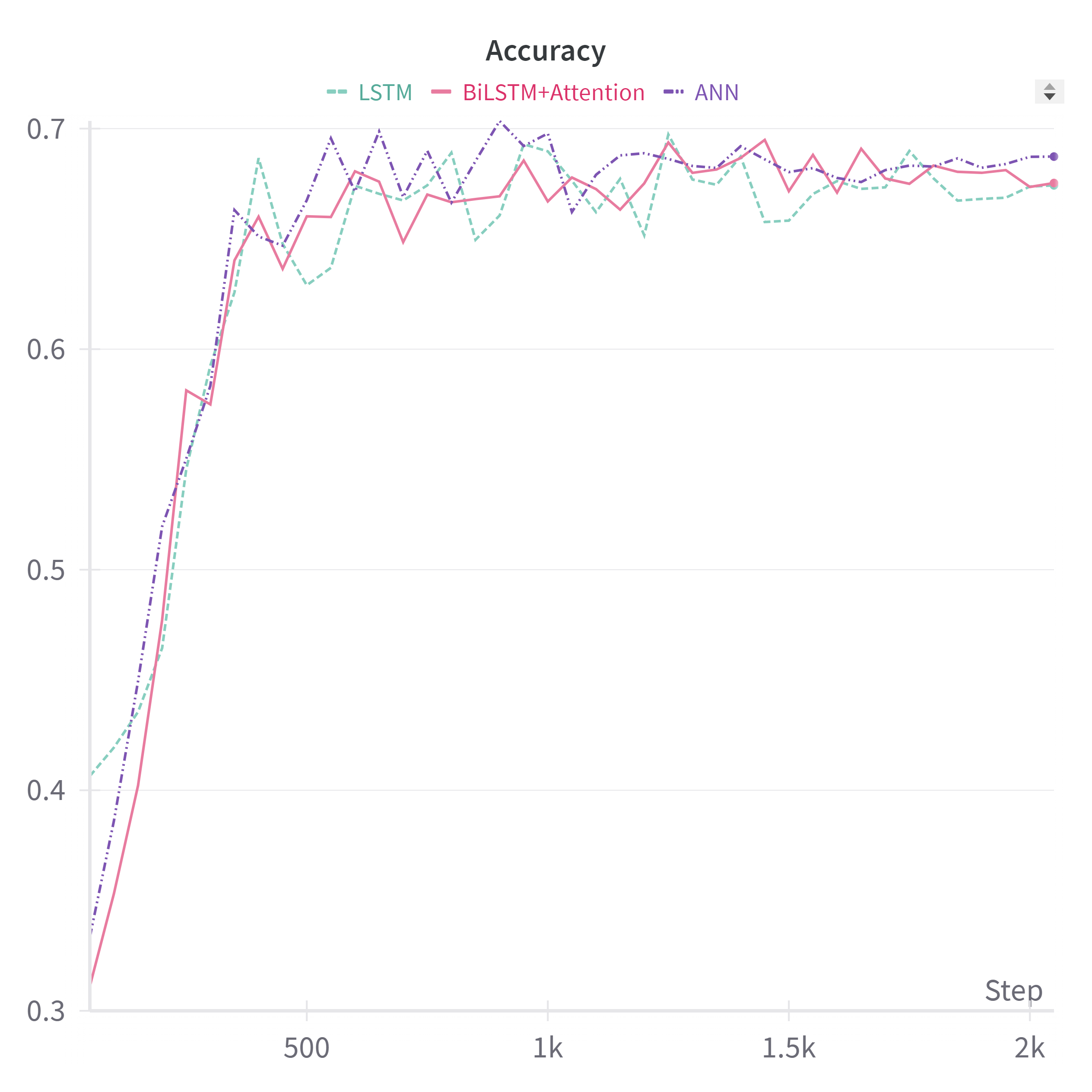}}
    \end{minipage}%
    \hfill
    \begin{minipage}{.27\textwidth}
        \centering
        \fbox{\includegraphics[scale=0.06]{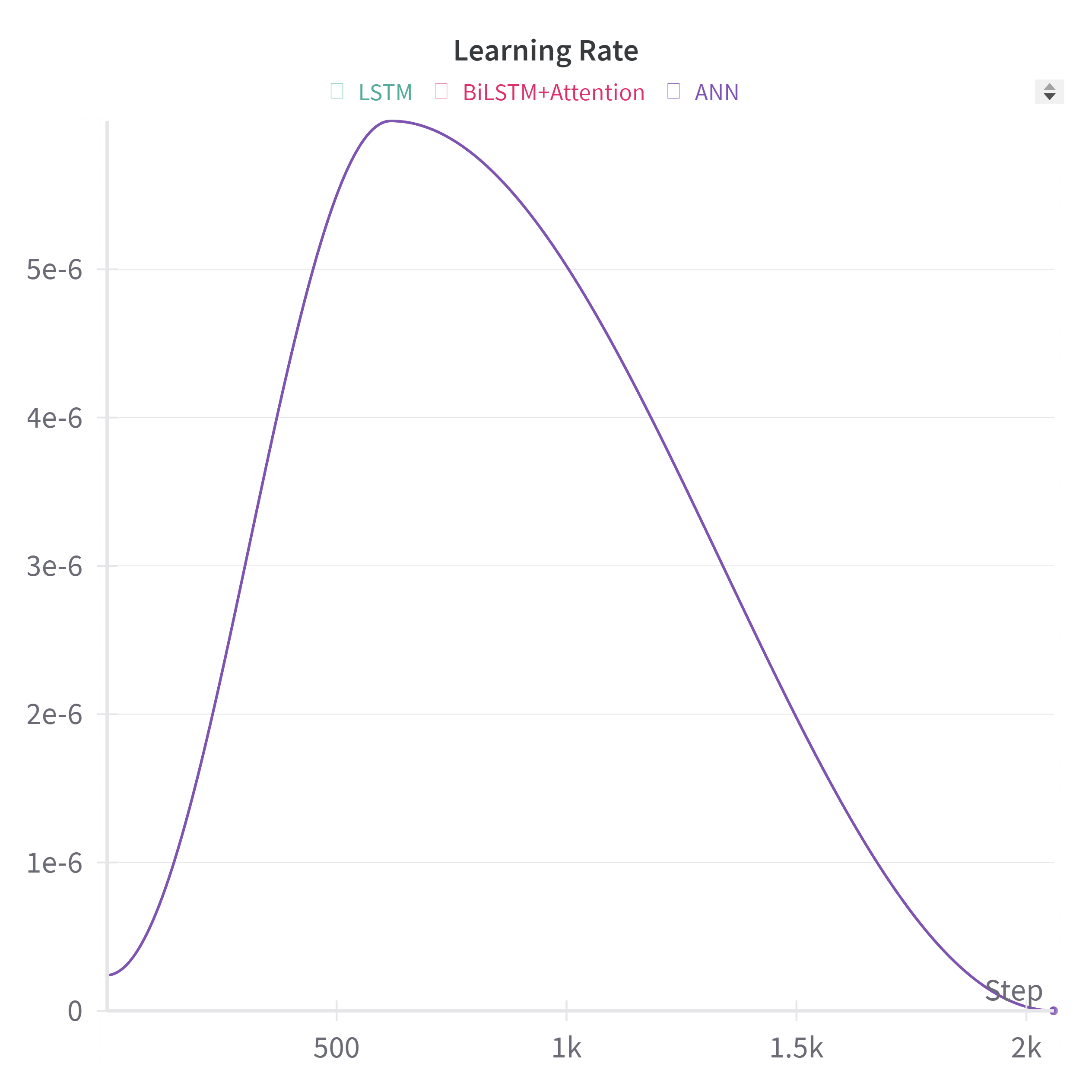}}
    \end{minipage}%
    \hfill
    \begin{minipage}{.27\textwidth}
        \centering
        \fbox{\includegraphics[scale=0.06]{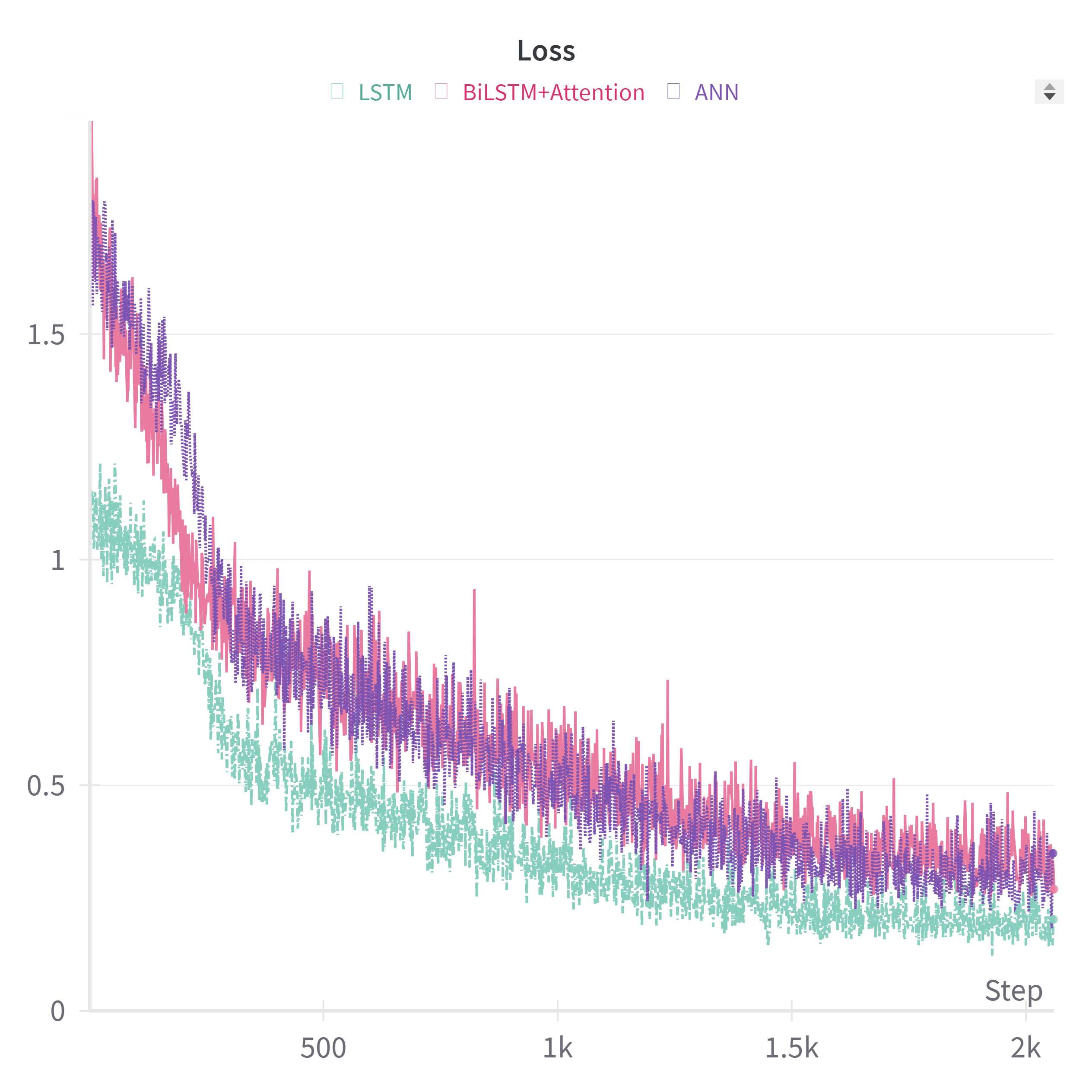}}
    \end{minipage}
    \caption{Accuracy, Learning rate and Loss of the sentiment model}
    \label{fig:combined}
\end{figure}
Due to parts of the dataset being manually annotated, I was able to achieve a 90\% accuracy rate of manual annotation using the gpt3.5 API; for details of this part's prompt, please see the appendix. After using the gpt3.5 API to annotate 30,000 pieces of data, my local language model achieved about 70\% accuracy of the gpt3.5 annotations. This model is based on the transformer framework, with a custom classifier head added on top. The `TextDataset` class I adopted normalizes the text (converting it to lowercase) and performs tokenization, utilizing the tokenizer from the RoBerta Large pre-trained model.
\\
After annotating each piece of news with custom sentiment, I started building sentiment factors for GAS. I did two things:
\begin{enumerate}
    \item For each date in the dataset, it calculates the proportion of positive and negative sentiments over the past 7 days and 30 days, including that day.
    \item This proportion is obtained by dividing the number of positive (or negative) sentiments in a specific time window by the total number of sentiments (positive, neutral, negative) in the same time window.
\end{enumerate}
The formulas for these proportions are as follows:
\begin{itemize}
    \item \textbf{Positive Ratio (7d)}: \( \text{Positive Ratio (7d)} = \frac{\sum_{i=t-6}^{t} P_i}{\sum_{i=t-6}^{t} T_i} \) where `t` represents the current date.
    \item \textbf{Positive Ratio (30d)}: \( \text{Positive Ratio (30d)} = \frac{\sum_{i=t-29}^{t} P_i}{\sum_{i=t-29}^{t} T_i} \)
    \item \textbf{Negative Ratio (7d)}: \( \text{Negative Ratio (7d)} = \frac{\sum_{i=t-6}^{t} N_i}{\sum_{i=t-6}^{t} T_i} \)
    \item \textbf{Negative Ratio (30d)}: \( \text{Negative Ratio (30d)} = \frac{\sum_{i=t-29}^{t} N_i}{\sum_{i=t-29}^{t} T_i} \)
\end{itemize}
where P represents the number of positive sentiments, N represents the number of negative sentiments, and T represents the total number of sentiments (i.e., P + N + neutral sentiments).

\subsubsection{Feature Transformation}

\begin{itemize}
    \item Use the \texttt{train\_test\_split} method to divide the data into a training set and a test set. The test set accounts for 20\% of the total data, and this step does not involve shuffling of data (\texttt{shuffle=False}) to maintain the time series nature of the data.
    \item Further, divide the training set into a smaller training set and a validation set. The validation set accounts for 10\% of the original training set, and again, no shuffling is performed.
    \item Initialize \texttt{StandardScaler}.
    \item Use the \texttt{fit\_transform} method on the training set. This step first calculates the mean and standard deviation (\texttt{fit}) of the training set, and then standardizes the training set data (\texttt{transform}) using these parameters.
    \item Use the \texttt{transform} method on the validation set and test set. It is important here to use the mean and standard deviation of the training set, not those of the validation or test sets. This ensures that the model does not encounter any information from the test or validation sets during training, maintaining consistency in data processing.
\end{itemize}

\subsubsection{Feature Selection} 
Feature selection is used to reduce complexity and solve multicollinearity. Due to the relationship between data volume and features, I need to control the number of features.

\paragraph{Filter Methods} 
Filter-based methods evaluate the importance of features by calculating a ranking criterion, which then helps to determine correlation scores. Subsequently, features with low scores that are either below a predefined threshold or exceed a specific count are removed from the dataset. In this framework, the Pearson correlation method is included as it excels in reflecting the impact of continuous variables, while the Chi-square method is more adept at capturing the impact of discrete variables.

\begin{itemize}
    \item \textbf{SelectPercentile:} I use SelectPercentile (\texttt{f\_classif}, percentile) for feature selection. \texttt{f\_classif} is used to determine the scoring criteria for feature selection, and percentile is used to determine the proportion of feature selection. This helps us roughly determine the range of parameter numbers.
    \begin{figure}[H]
        \centering
        \fbox{\includegraphics[scale=0.33]{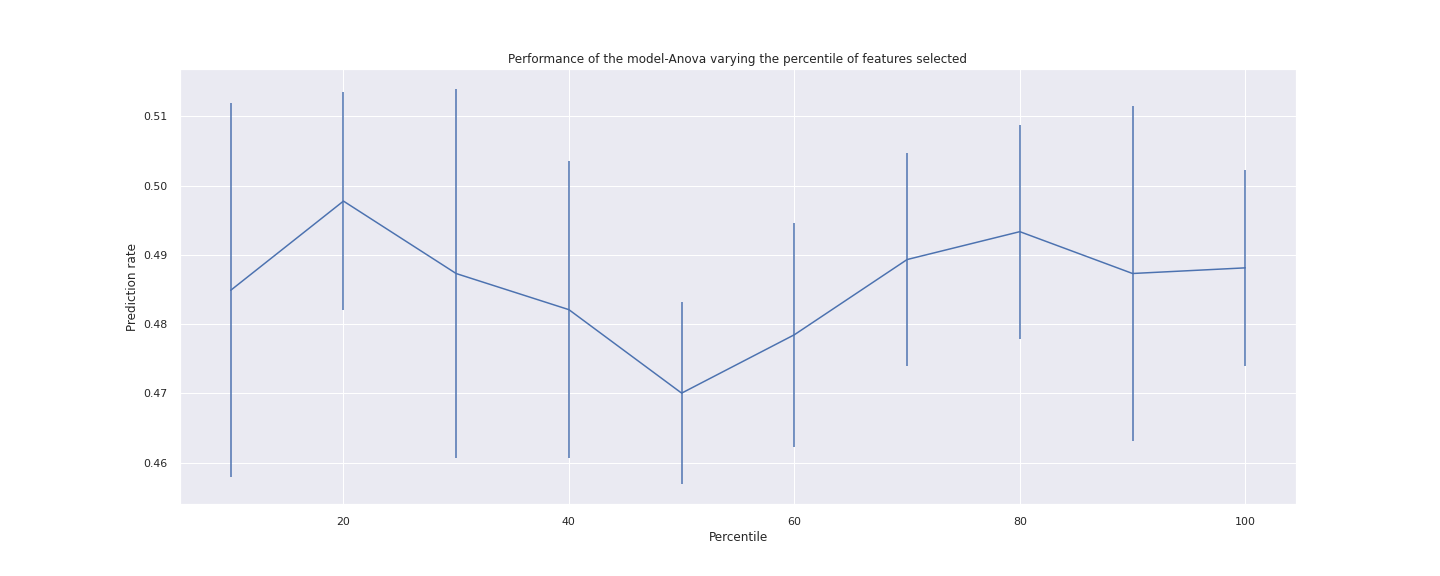}} 
        \caption{factors distribution by Anova}
    \end{figure}
    As seen in the figure, the best performance is achieved when the number of factors is around 30.

    \item \textbf{Variance Inflation Factor (VIF):} The variance inflation factor (VIF) was employed to provide a measure of multicollinearity among the independent variables in a multiple regression model, as detecting multicollinearity is crucial. Although it does not reduce the explanatory power of the model, it may decrease the statistical significance of the independent variables. A VIF greater than 5 indicates a high degree of multicollinearity, and I selected 151 factors with VIF less than 5.

    \item \textbf{Chi-Square:} Chi-Square was also taken into account to determine whether the association between two categorical variables in the sample reflects their real association in the population, based on their frequency distribution. By calculating the Chi-square statistic for each variable in relation to the target variable, the existence and strength of their relationship are determined. Subsequently, variables with weaker relationships to the target variable are excluded from consideration.

    \item \textbf{Pearson Correlation:} Pearson correlation involves using the covariance matrix between variables to calculate correlation coefficients. In this framework, the Pearson correlation method is included as it excels in reflecting the impact of continuous variables, while the Chi-square method is more adept at capturing the impact of discrete variables.
    \begin{figure}[H]
    \centering
    \fbox{\includegraphics[scale=0.35]{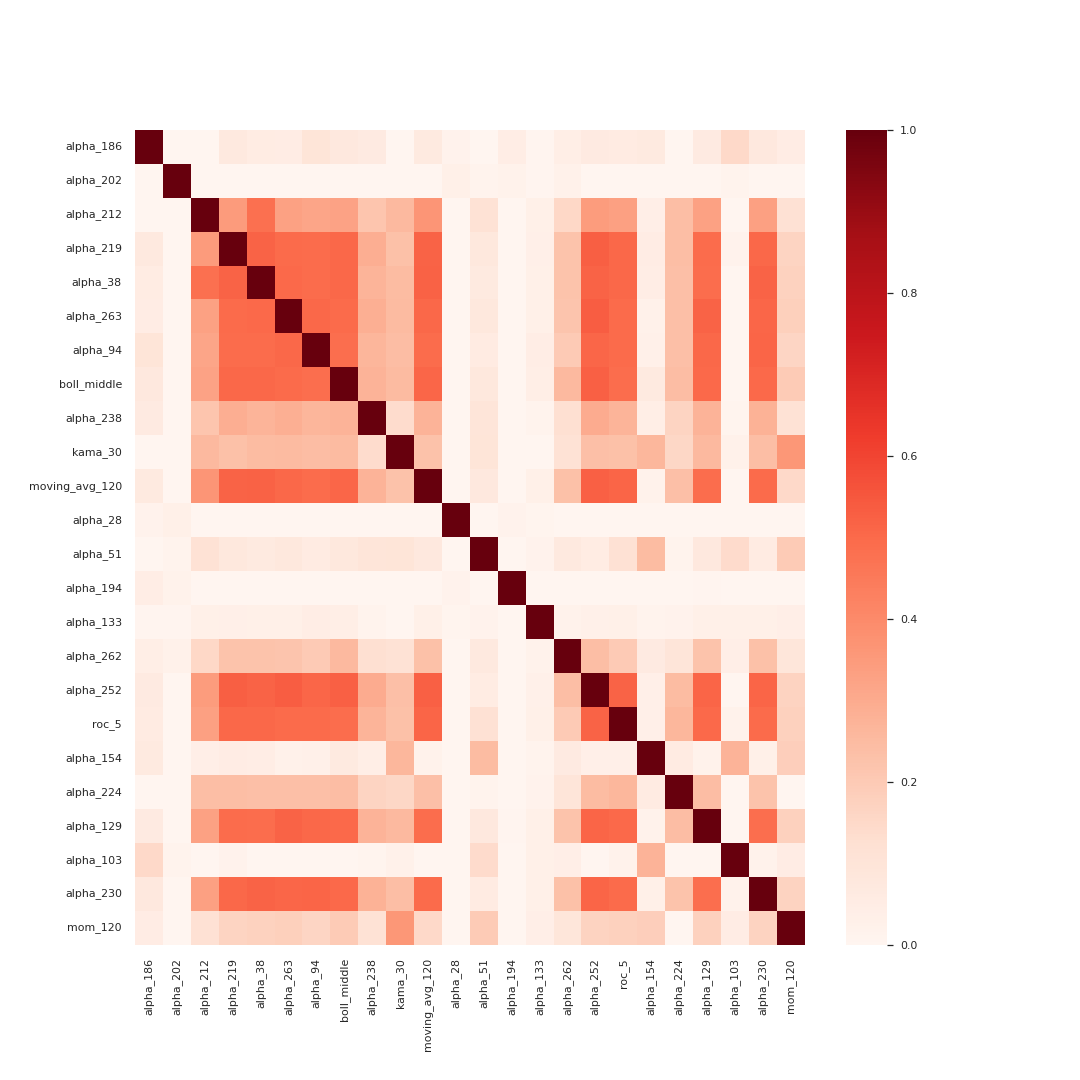}} 
    \caption{Factors Correlation}
    \end{figure}
    It is clear see that the correlation of the selected factors is not significant.
\end{itemize}

\paragraph{Wrapper Method}
\begin{itemize}
    \item\textbf{Boruta Algorithm}
    The Boruta algorithm is a feature selection method, fundamentally based on two concepts: shadow features and binomial distribution. This algorithm can automatically perform feature selection on datasets. The Boruta function evaluates the importance of each variable in a cyclical manner, comparing the importance of the original variables with that of the shadow variables in each iteration. An original variable is considered important if its importance significantly exceeds that of the shadow variables; conversely, it is deemed unimportant if its importance is significantly lower than that of the shadow variables.
    
    \begin{itemize}
        \item Starting with a dataset \( X \), for each real feature \( R \), the order is randomly shuffled, and these shuffled original features are termed as shadow features. At this point, a shadow dataframe is appended to the original dataframe, resulting in a new dataframe whose number of columns is twice that of \( X \).
        \item Then, it trains an extended dataset with a random forest classifier, using a feature importance measure (default set to mean decrease in accuracy) to evaluate the importance of each feature. The higher the importance, the more significant the feature.
        \item In each iteration, it checks whether a real feature has higher importance than the best shadow feature (i.e., whether the feature has a higher score than the highest shadow feature) and continuously removes those deemed as very unimportant.
        \item Finally, the algorithm stops when all features are either confirmed or rejected, or when a predetermined limit of random forest runs is reached.
    \end{itemize}
    
    Boruta follows all relevant feature selection methods, capturing all features related to the outcome variable. This method minimizes the error of the random forest model, ultimately forming an optimized subset of minimal features. By selecting a highly streamlined version of an input dataset, it may result in the loss of some relevant features. Boruta does not address the issue of feature collinearity, hence I used filter methods.

    \item\textbf{SHAP}
    provide an alternative method for assessing the importance of feature permutations by reducing model performance. On the other hand, SHAP scores depend on the size of feature attribution. Shapley value is the average marginal contribution of all potential alliances divided by the total number of alliances. SHAP values clarify the impact of features on predictions rather than their impact on model fitting, providing a better understanding of variable contributions. To evaluate this difference and its impact on predictive performance, SHAP values are incorporated into the framework.
    \begin{figure}[H]
    \centering
    \fbox{\includegraphics[scale=0.6]{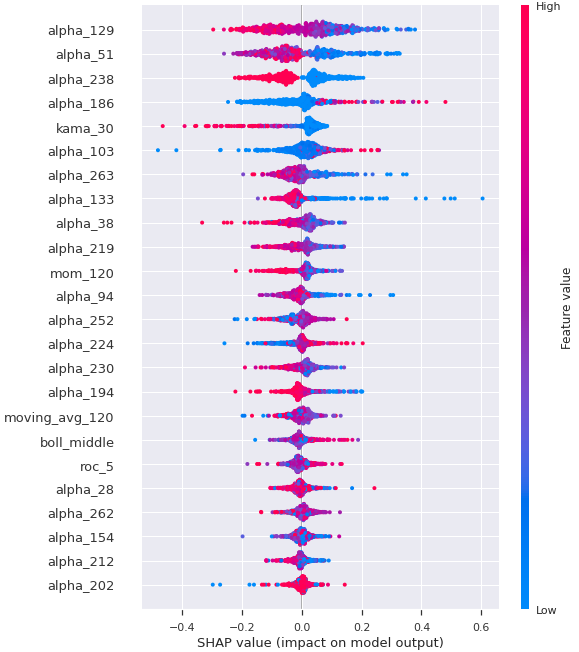}} 
    \caption{factors shap value}
    \end{figure}
    \end{itemize}
 
I selected 34 factors through feature selection. Based on these factors and the IC value of return, I conducted a buy-and-hold strategy for Bitcoin. The results of single-factor analysis show that there are performance differences among the factors.
\begin{figure}[H]
\centering
\fbox{\includegraphics[scale=0.5]{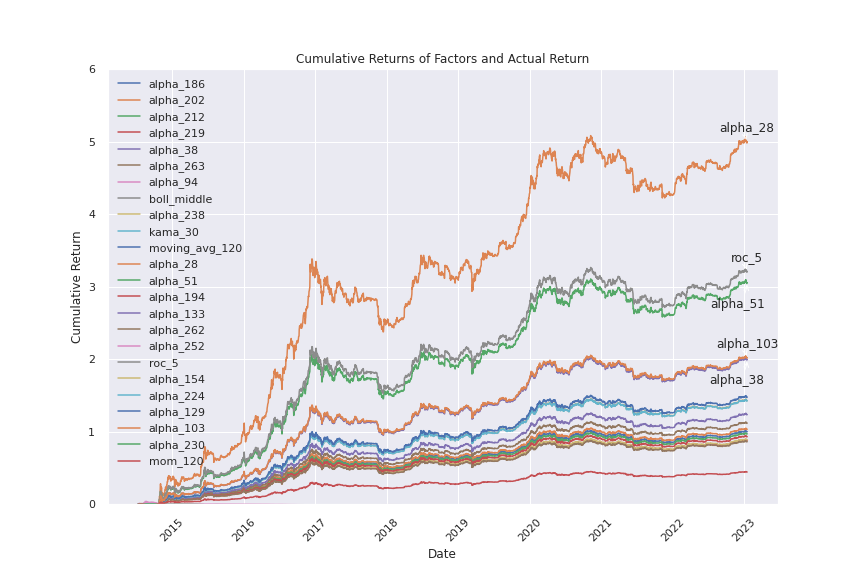}} 
\caption{cumulative return by factors}
\end{figure}

\subsection{Model Architecture}
\begin{figure}[H]
    \centering
    \fbox{\includegraphics[scale=0.4]{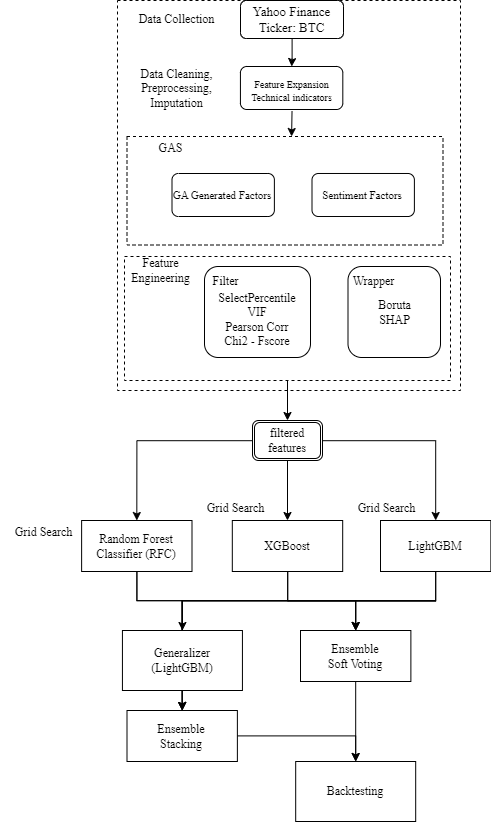}} 
    \caption{Model Architecture}
\end{figure}

\subsubsection{Model Selection}
This paper uses three base learners to build the blending ensemble model.

\paragraph{RandomForest}
RandomForest is essentially an implementation of Bagging on decision trees, where Bagging is an effective method to reduce the variance of a learner's results. The principle involves generating more varied samples through bootstrap, learning with a learner on each sample, and finally determining the result by voting. RandomForest adds a unique feature: in addition to random sampling, it also limits the number of features during bootstrap sampling, enriching the variety of trees in the forest.
\paragraph{XGBoost}
Gradient Boosting, a type of Boosting method, builds its model in a step-by-step manner by reducing the residuals of the previous learner, thus establishing a new model in the direction of reduced residuals. Unlike the initial boosting methods that amplify the importance of wrongly classified data to make the new learning set focus more on wrongly classified points, Gradient Boosting optimizes the algorithm step by step in the direction of reducing residuals. XGBoost is an improvement over Gradient Boosting, with the following enhancements:
\begin{enumerate}
    \item It uses a second-order Taylor expansion to find the gradient, which is more efficient and yields better results than the previous first-order approach.
    \item In terms of leaf nodes, XGBoost not only limits the total number of leaf nodes but also incorporates an L2 regularization term to prevent overfitting.
\end{enumerate}
\paragraph{LightGBM}
LightGBM, a boosting package developed by Microsoft, is one of the best-performing boosting algorithms currently available. Similar to XGBoost, it shifts from depth constraint (level-wise) to leaf constraint (leaf-wise), significantly enhancing both speed and accuracy.

\paragraph{Stacking Method}
Stacking, also known as Stacked Generalization is an ensemble technique that combines multiple classifications or regression models via a meta-classifier or a meta-regressor. The base-level models are trained on a complete training set, then the meta-model is trained on the features that are outputs of the base-level model. The base-level often consists of different learning algorithms and therefore stacking ensembles are often heterogeneous. This paper uses LGBM as the base model.

\paragraph{Soft Voting Method}
Soft Voting: In soft voting, the output class is the prediction based on the average of probability given to that class. 
In practical the output accuracy will be more for soft voting as it is the average probability of the all estimators combined. That's why I didn't use hard voting this time. And the soft voting can be a benchmark for comparing the performance of stacking method.

\subsubsection{Model Training}
For hyperparameters, all base learners uses grid search for optimization. The initial value of these three base learners were taken from \citep{hasan2022blending} research.
\begin{table}[h]
\begin{adjustbox}{width=300pt,center}
\begin{tabular}{ll}
\hline
Model & Values of Hyperparameters \\ \hline
XGBoost (XGB) & $\alpha=0.3$, $lr=0.1$, $colsample_bylevel=0.4$\\
Random Forest (RF) & $n_{\text{estimators}}=500$, $random_{\text{state}}=42$ \\
LightGBM (LGBM) & Learning Rate $=0.1$, $num_{\text{leaves}}=64$ \\
\hline
\end{tabular}
\end{adjustbox}
\end{table}

Below is a snapshot of the grid search parameters of Random Forest Classifier, because it spends so much time, I am not able to get one for XGBoost and LightGBM.
\begin{figure}[H]
    \centering
    \fbox{\includegraphics[scale=0.35]{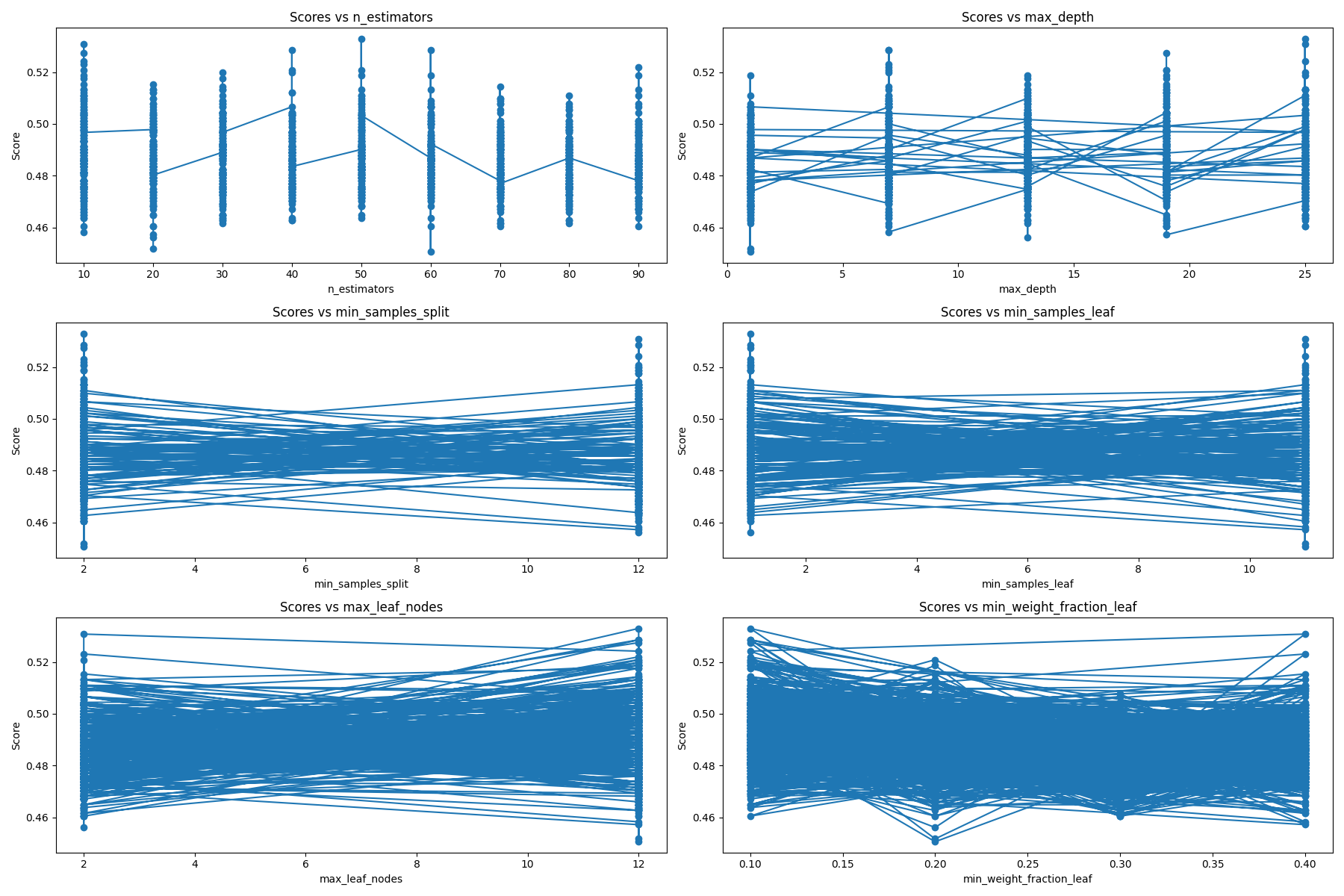}} 
    \caption{Grid Search Parameters for rfc}
    \label{fig:mc_ver}
\end{figure}
The Random Forest model's tuning encompassed parameters such as n\_estimators (ranging from 10 to 100 in steps of 10), max depth (from 1 to 28 in steps of 2), min samples split and leaf (ranging from 2 to 20 and 1 to 20 in steps of 2, respectively), max leaf nodes (from 2 to 20 in steps of 2), and min weight fraction leaf (from 0.1 to 0.4 in steps of 0.1), with a fixed random state of 42. 

For the XGBoost model, the parameters tuned included learning rate (set to 0.1, 0.3, 0.5), colsample by level and by tree (both set to 0.4), max depth (2, 3), min child weight (0.05, 3, 5), lambda (0.05, 0.3), alpha (0.05, 0.3), and gamma (0.05, 0.3). 

For LightGBM, I focused on parameters including the objective set to 'binary', boosting type as 'gbdt', n\_estimators at 1000, learning rate at 0.05, num leaves at 64, max depth set to 3, colsample bytree at 0.65, reg alpha at 1.2, and reg lambda at 1.4. These parameters were meticulously chosen to explore a comprehensive range of configurations, aiming to optimize each model's performance.
\\
All the parameters will updated for the original classifier after the grid search.

\subsection{Model Evaluation \& Results} \label{sec:result}
Results show below:
\begin{figure}[h]
    \centering
    \begin{minipage}{.27\textwidth}
        \centering
        \fbox{\includegraphics[scale=0.35]{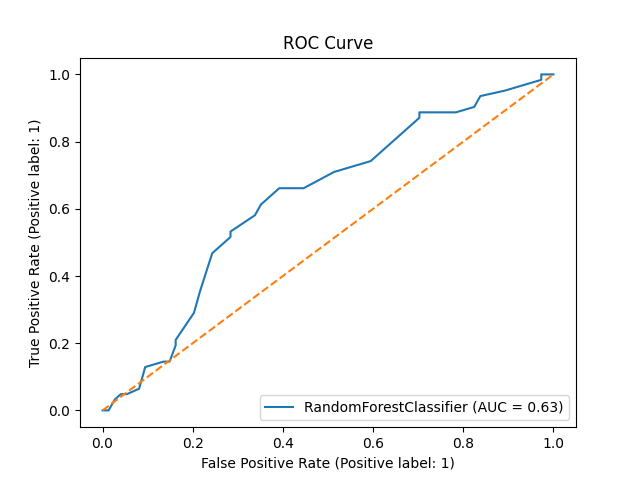}}
    \end{minipage}%
    \hfill
    \begin{minipage}{.27\textwidth}
        \centering
        \fbox{\includegraphics[scale=0.35]{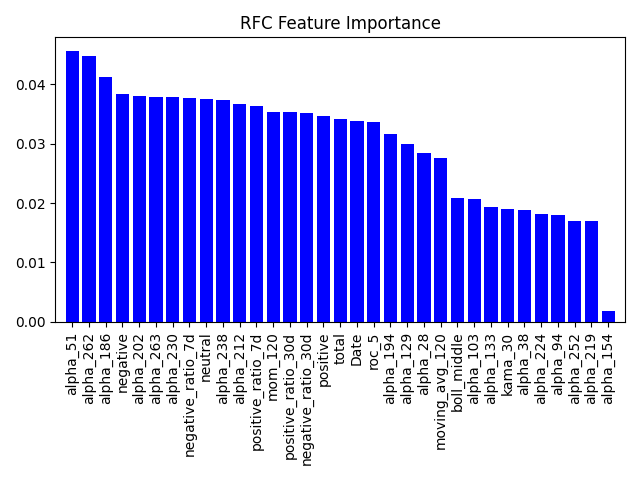}}
    \end{minipage}%
    \hfill
    \begin{minipage}{.27\textwidth}
        \centering
        \fbox{\includegraphics[scale=0.35]{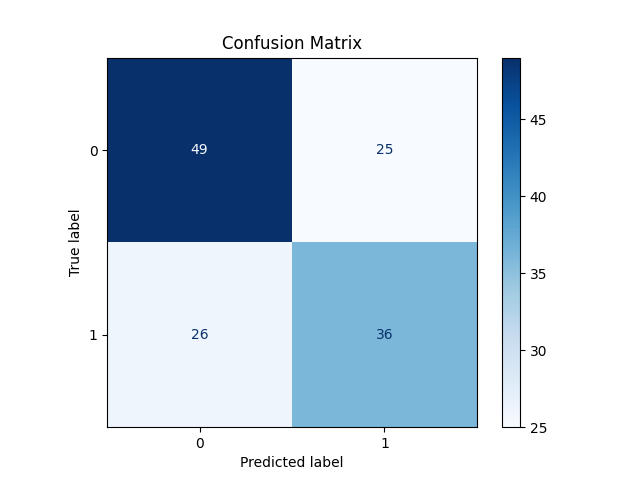}}
    \end{minipage}
    \caption{The Prediction Quality of RFC}
    \label{fig:combined}
\end{figure}

\begin{figure}[h]
    \centering
    \begin{minipage}{.27\textwidth}
        \centering
        \fbox{\includegraphics[scale=0.35]{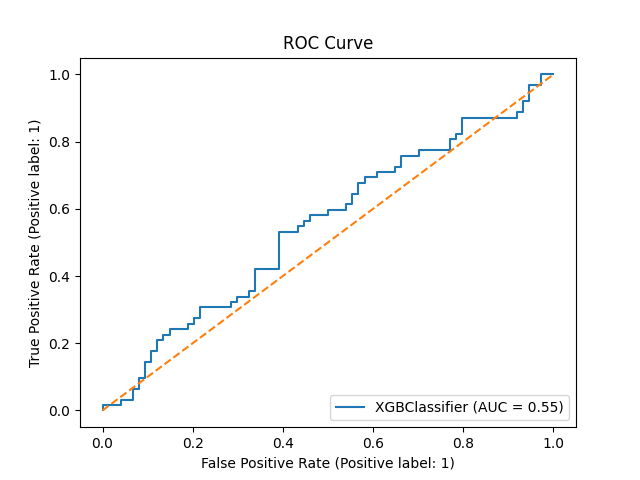}}
    \end{minipage}%
    \hfill
    \begin{minipage}{.27\textwidth}
        \centering
        \fbox{\includegraphics[scale=0.35]{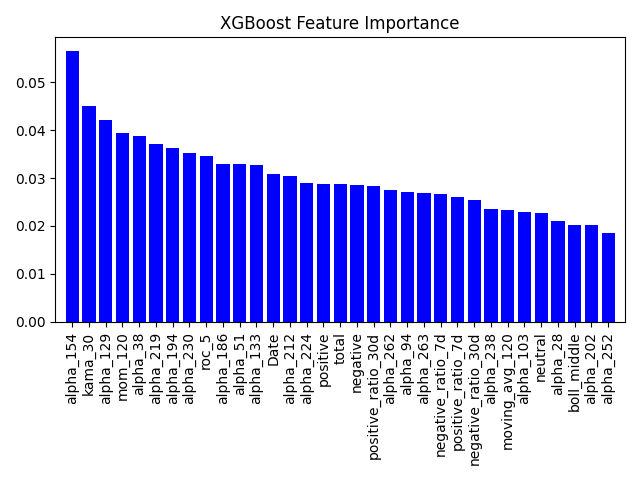}}
    \end{minipage}%
    \hfill
    \begin{minipage}{.27\textwidth}
        \centering
        \fbox{\includegraphics[scale=0.35]{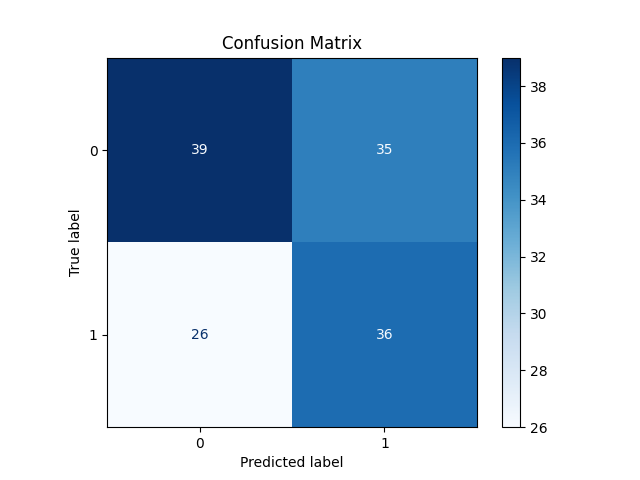}}
    \end{minipage}
    \caption{The Prediction Quality of XGBoost}
    \label{fig:combined}
\end{figure}

\begin{figure}[h]
    \centering
    \begin{minipage}{.27\textwidth}
        \centering
        \fbox{\includegraphics[scale=0.35]{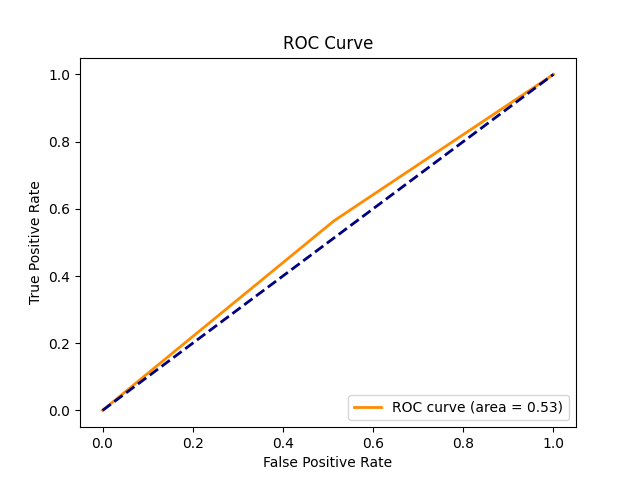}}
    \end{minipage}%
    \hfill
    \begin{minipage}{.27\textwidth}
        \centering
        \fbox{\includegraphics[scale=0.35]{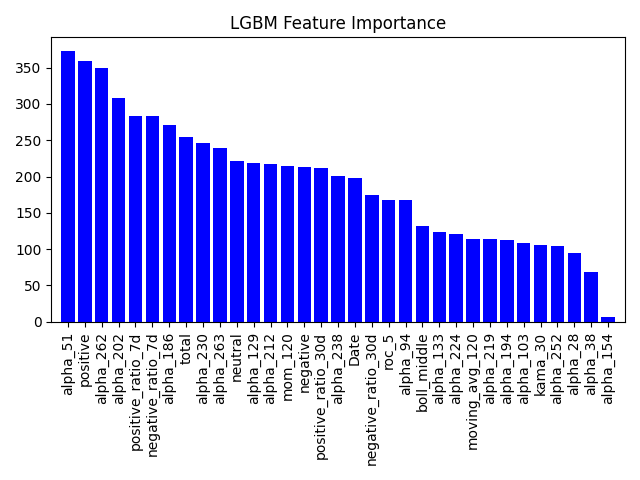}}
    \end{minipage}%
    \hfill
    \begin{minipage}{.27\textwidth}
        \centering
        \fbox{\includegraphics[scale=0.35]{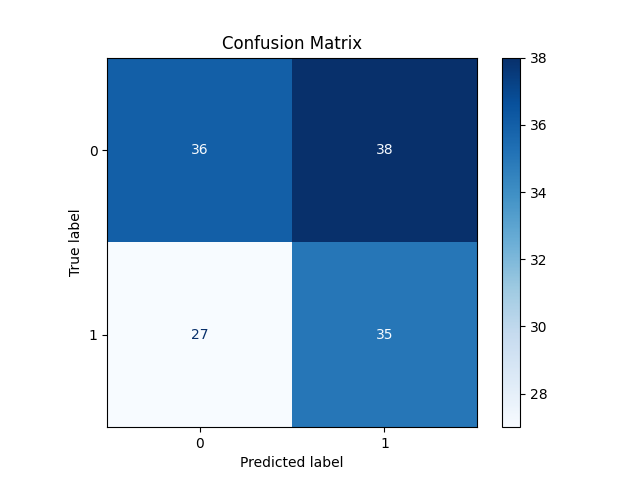}}
    \end{minipage}
    \caption{The Prediction Quality of LightGBM}
    \label{fig:combined}
\end{figure}

\begin{table}[H]
\centering
\caption{Classification Table of LightGBM, Random Forest Classifier (RFC), and XGBoost}
\label{tab:classifier_metrics}
\begin{tabular}{lcccccc}
\hline
\multicolumn{1}{c}{} & \multicolumn{2}{c}{\textbf{LightGBM}} & \multicolumn{2}{c}{\textbf{RFC}} & \multicolumn{2}{c}{\textbf{XGBoost}} \\ 
\cline{2-7}
\textbf{Metric}      & 0             & 1            & 0         & 1        & 0          & 1           \\ 
\hline
Precision            & 0.57          & 0.48         & 0.59      & 0.52     & 0.60       & 0.51        \\
Recall               & 0.49          & 0.56         & 0.65      & 0.45     & 0.53       & 0.58        \\
F1-Score             & 0.53          & 0.52         & 0.62      & 0.48     & 0.56       & 0.54        \\ 
\hline
\end{tabular}
\end{table}

I made a prediction for each classifier, as well as a prediction for the voting classifier and stacking model. And I used the following strategy: if the prediction is positive, buy Bitcoin. Otherwise, sell short. The results are as follows:
\begin{figure}[H]
    \centering
    \fbox{\includegraphics[scale=0.5]{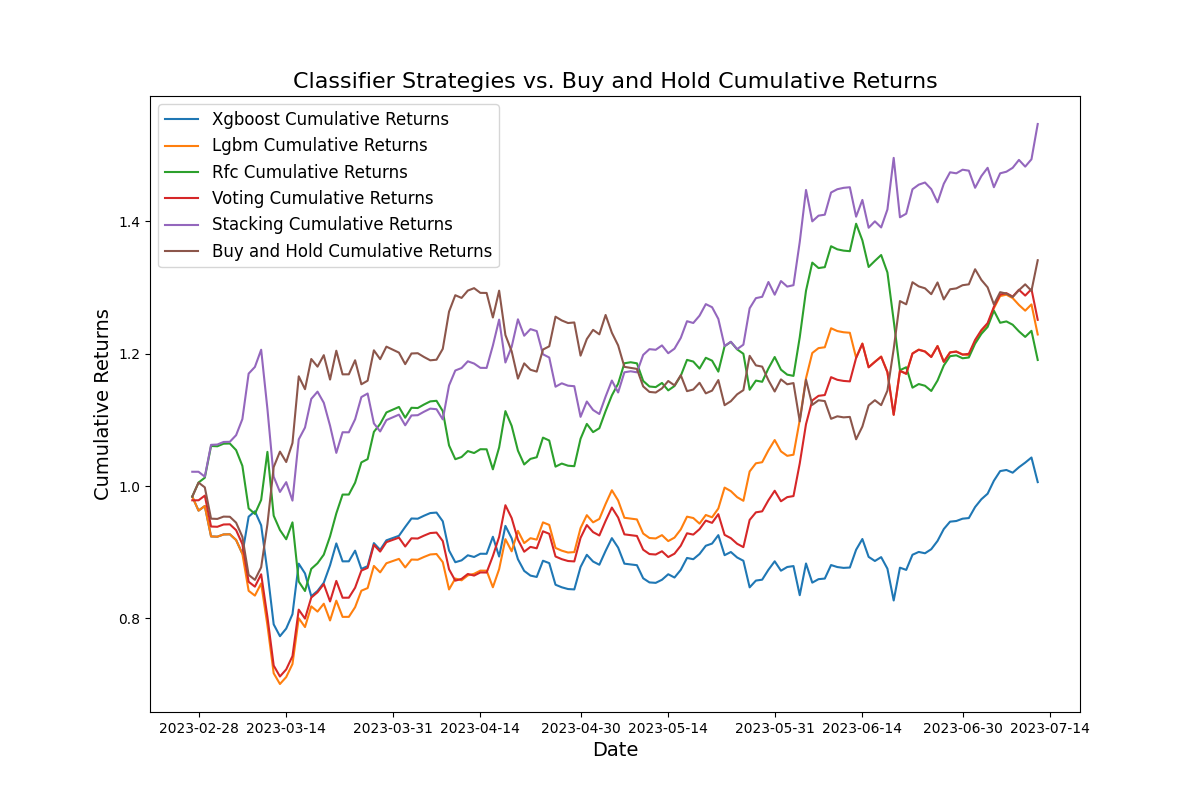}} 
    \caption{Backtesting Result}
\end{figure}
The Stacking Model has demonstrated its strength.

\section{Observations \& Discussions} \label{sec:obs}
In the comparative analysis of the LightGBM, Random Forest Classifier (RFC), and XGBoost models, distinct performance patterns emerge. The RFC exhibits a slightly superior balance between precision and recall for class 0, evidenced by a precision of 0.59 and recall of 0.65, leading to the highest F1-score of 0.62 in this class. This suggests that RFC is relatively more effective in correctly identifying class 0 instances. 

In contrast, LightGBM and XGBoost demonstrate a more balanced performance across both classes, with LightGBM showing a marginal preference towards class 1 (recall of 0.56) and XGBoost favoring class 1 with a higher recall of 0.58. However, both models exhibit a lower overall accuracy compared to RFC, as indicated by their respective accuracy scores. 

It is noteworthy that XGBoost achieves a slightly higher precision for class 0 (0.60) but does not translate this advantage into a correspondingly high F1-score, which is indicative of its challenges in balancing recall and precision. The consistency in the macro and weighted averages across all models suggests a relatively uniform performance across both classes, although the subtle differences in precision and recall reflect their unique strengths and limitations in handling class-specific characteristics.
\\
\\
In addition, I found that alpha51 and alpha262 are both important for LGBM and RFC, while 238 is a factor that appears in all three classifiers. Regarding 238 and 262, I can provide some explanations. For example, the 238 factor calculates the time series argmin (index of minimum value) of accumulated logarithmic returns within a period of 40 days. Essentially, it identifies points where the cumulative logarithmic return in the time series reaches its lowest value over the past 40 days. This can identify potential turning points or periods of relative underperformance in asset returns. However, it is difficult to explain factors like alpha51. This also demonstrates the effectiveness of GAS in complex high-dimensional operations.

\begin{table}[h]
\centering
\caption{The best alpha expressions}
\begin{tabular}{lp{0.7\textwidth}}
\hline
Factor & Formula \\ \hline
factor\_238 & $ts\_argmin(CumLgReturn\_40d)$ \\
factor\_262 & $gp\_mul(roc\_30, roc\_5)$ \\
factor\_51 & $delta(gp\_add(gp\_sqrt(gp\_max(ts\_argmin(gp\_tan(ts\_sum(cci\_cci))),$
$gp\_abs(gp\_sin(gp\_sin(CumLgReturn\_110d))))),$
$rank(delta(gp\_mul(gp\_mul(gp\_cos(CumLgReturn\_3d),$
$ts\_argmax(CumLgReturn\_140d)),$
$gp\_sub(gp\_mul(cci\_cci, moving\_avg\_20),$
$scale(moving\_avg\_120))))))$ \\
\hline
\end{tabular}
\end{table}

In addition, I should consider optimizing the model. For event-driven assets like Bitcoin, I should separate and process training data from different periods. \cite{huang2009hybrid} combined Support Vector Regression (SVR) with Self-Organizing Feature Map (SOFM) technology and filter-based feature selection to reduce training time costs and improve prediction accuracy. In the first stage, SOFM clusters the training data into several disjoint clusters, each containing similar objects. In the second stage, a separate SVR model is built for each cluster to increase average prediction accuracy and potentially bring capital gains. Additionally, the SOFM-SVR model reduces training time by dividing the training samples into small sample-sized clusters that save a lot of time since SVR's training time depends on sample size. The periodicity of Bitcoin-related news also proves this necessity.
% 222222222222222222
\begin{figure}[H]
    \centering
    \fbox{\includegraphics[scale=0.4]{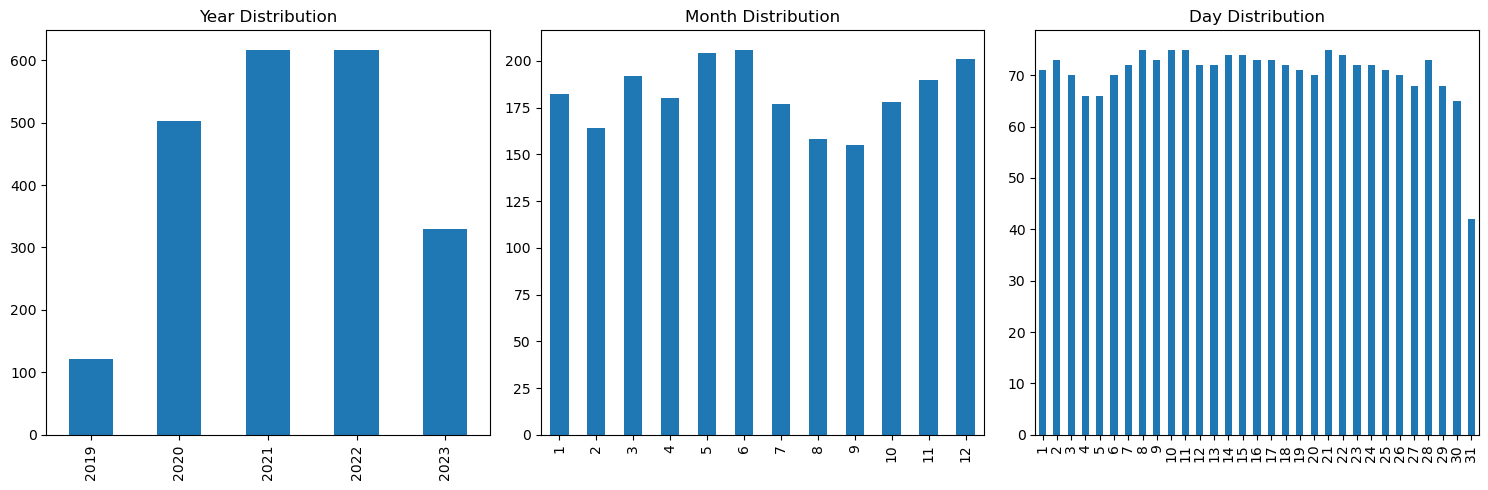}} 
    \caption{news records distribution}
    \label{fig:mc_ver}
\end{figure}

\section{Conclusion} \label{sec:conclusion}
This article proposes a Genetic-algorithm generated Alpha factors Sentiment (GAS) blending ensemble model aimed at predicting Bitcoin trends by integrating ensemble learning and feature selection algorithms. The GAS framework consists of thirty-four alpha factors and eight economic sentiment factors from news data, effectively capturing the daily trends of Bitcoin.

The model architecture combines the applications of LightGBM, XGBoost, and Random Forest Classifier (RFC) in stacked ensembles, demonstrating significant effectiveness in capturing the complexity of the Bitcoin market. The model utilizes complex feature engineering including genetic algorithm-generated alpha factors and sentiment analysis, combined with diverse data sources to identify key Alpha factors such as alpha51, alpha238, and alpha262 highlighting GAS's effectiveness in handling complex high-dimensional operations. Feature selection processes using methods such as SHAP and Boruta further optimize the model input enhancing its predictive ability.

Experimental results show that this model performs robustly on traditional buy-and-hold strategies with stacking method being particularly effective. Diversified performance across different indicators by individual classifiers (LightGBM, RFC, XGBoost) highlights the value of hybrid methods in capturing different aspects of market behavior. 

However, research also highlights a need for further optimization. Considering Bitcoin's sensitivity to events adopting more segmented data processing methods is desirable. Similar to SOFM-SVR models involving dividing training data into several non-overlapping clusters establishing separate models for each cluster can improve prediction accuracy while reducing training time costs. Combining attention to periodicity related to Bitcoin-related news events can produce more refined and effective predictions.

\singlespacing
\setlength\bibsep{0pt}

\section*{Appendix} \label{sec:appendixa}
\addcontentsline{toc}{section}{Appendix}
\begin{codelisting}[H]
\begin{lstlisting}[language=Python]
prompt = f"""
Your task is to analyze and report the overall sentiment of only one out of 
the following sentiments:
[positive, negative].
On the text delimited by triple backticks:

Taking into account the global macro environment, assess the impact of this 
news on the global economy or the countries mentioned in the text. 
Your analysis should be rational and Objective.

And these are four points you must obey:
1. Your output is solely composed of the tags I have given you. 
2. Output one of these two tags regardless of the situation.
3. you must output only one tag and strictly adhere to this requirement.
4. Your output can only fall within the binary tags: [positive, negative].

You must strictly follows:
Your output must either be the word "positive" or the word "negative",
it cannot be any other answer.
if you cannot decide, use the word "neutral".

When you cannot make a decision, try use these:
1. **Monetary Policy and Interest Rates**: 
Tight policy raises rates to control inflation, often dampening 
market sentiment.
2. **Interest Rates and Inflation Expectations**: 
Higher long-term rates with rising inflation expectations can negatively 
impact market mood.
3. **Phillips Curve**: Low rates reduce unemployment 
but increase inflation risk, impacting market sentiment ambiguously.
4. **International Capital Flows**: High rates attract foreign investment, 
affecting currency value and market sentiment variably.
5. **Investor Sentiment and Market Behavior**: 
Rising rates typically depress bond and stock markets, and vice versa.
6. **Keynes's Liquidity Preference Theory**: 
Extremely low rates can lead to a liquidity trap, negatively affecting 
market sentiment.
7. **Interest Rates and Economic Cycles**: 
Central banks adjust rates in response to economic cycles, influencing 
market mood.
Text to analyze: ```{comment}```
"""
\end{lstlisting}
\end{codelisting}
\label{code:code1}

\vspace{5em}
\bibliographystyle{apalike}
\bibliography{ref}

\end{document}